\documentclass[aps,floats,superscriptaddress,twocolumn,pre,showpacs,reprint]{revtex4}
\usepackage{amssymb,amsmath}
\usepackage{amsmath,amssymb}
\usepackage{graphicx}
\usepackage{psfrag,textcomp}

\def\beq{\begin{equation}}
\def\eeq{\end{equation}}
\def\bea{\begin{eqnarray}}
\def\eea{\end{eqnarray}}
\begin{document}
\title{Generic nonequilibrium steady states in asymmetric
exclusion processes on a ring with bottlenecks}

\author{Niladri Sarkar}\email{niladri.sarkar@saha.ac.in}
\author{Abhik Basu}\email{abhik.basu@saha.ac.in}
\affiliation{Condensed Matter Physics Division, Saha Institute of
Nuclear Physics, Calcutta 700064, India}
\date{\today}

\begin{abstract}
Generic inhomogeneous steady states in an asymmetric exclusion
process on a ring with a pair of point bottlenecks are studied.
We show that due to an underlying universal feature not
 considered hitherto, measurements of coarse-grained steady state densities in this model resolve the
 bottleneck structures only partially. Unexpectedly,
 it
 displays localization-delocalization transitions, and confinement of
delocalized domain walls, controlled by the interplay between
particle number conservation and
 bottleneck competition for moderate particle densities.


\end{abstract}

\maketitle
\section{Introduction}

Simplest physical modeling of classical transports in low dimensions
are often made in terms of asymmetric exclusion processes. For
instance, one-dimensional (1D) totally asymmetric simple exclusion
process (TASEP) with open boundaries provides a simple physical
description of restricted 1D motion in various natural~\cite{exam}
and social phenomena~\cite{traffic}; see Refs.~\cite{rev1,rev2} for
basic reviews on asymmetric exclusion processes.
In this article we investigate the generic relationship between the
inhomogeneous steady state densities, and conservation laws and
structural deformations in asymmetric exclusion processes with
periodic boundary conditions. To this end, we study the generic
inhomogeneous steady states in a 1D model that executes TASEP on a
ring having a pair of bottlenecks. We show that bottleneck
competitions in the model leads to {\em screening} or irrelevance of
one bottleneck by the other for moderate densities. Significantly,
this implies that coarse-grained measurements of the inhomogeneous
densities in closed TASEP cannot be reliably used to obtain
information about the underlying microscopic bottlenecks, as
experiments detect them only partially, establishing an underlying
universal feature distinctly different from critical phenomena or
critical dynamics. Furthermore, for moderate densities depending
upon the strengths of the bottlenecks, our model displays both
localized (LDW) and delocalized (DDW) domain walls, in contrast to
open TASEPs~\cite{openbottleneck}. Unexpectedly, DDWs can be {\em
smoothly confined} by tuning the relative positions of the
bottlenecks. Our results have experimental implications, e.g., in
studies of unidirectional circular ribosome translocations along
messenger RNA (mRNA) loops with defects or {\em slow codons} in
cells~\cite{chou,deftwo}.
Our model should serve as an important step for theoretical analysis
of the mutual interplay between particle number conservation and
arbitrary number of discrete bottlenecks in asymmetric exclusion
processes in 1D closed systems. The rest of the paper is organized
as follows: In Sec.~\ref{model}, we construct our model. Then in
Secs.~\ref{idphase} and \ref{ldhd}, we elucidate the different
inhomogeneous and homogeneous density phases of the system. Finally,
we summarize in Sec.~\ref{conclu}.

\section{Construction of our model}\label{model}

\begin{figure}[htb]
\includegraphics[height=4cm,width=5.0cm]{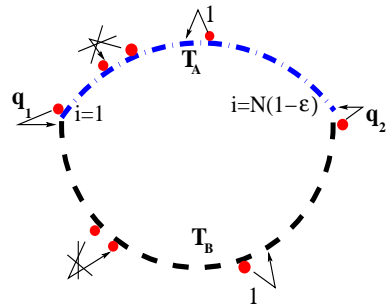}
\caption{(color online) Schematic diagram of our model; $T_A$ (blue dashed-dotted line) and
$T_B$ (black dashed line) are marked (see text).}
 \label{modeldiag}
\end{figure}

  Our 1D model consists of a ring having $2N$
sites, with two bottlenecks (point defects) of reduced hopping rates
$q_1,\,q_2<1$,  from $i=1$ to $  2N$ and
$i=N(1-\epsilon)+1$ to $ N(1-\epsilon), |\epsilon|\leq 1$, respectively.
Hopping rates elsewhere is unity; see
Fig.~(\ref{modeldiag}).
 Site labels $i$ run clockwise from $i=1$, whereas particles move anticlockwise.
   When one of
$q_1,\,q_2,|\epsilon|$ is set to unity, our model is physically
identical to that of Ref.~\cite{lebo}. It is convenient to use a
continuum labeling in thermodynamic limit (TL):
$N\rightarrow\infty$, $x=i/(2N), 0<x<1$. The bottleneck positions
are then at $x=0$ and $x=(1-\epsilon)/2$. With $N_p$ number of
particles in the system, we define a mean density $n=N_p/(2N)$. The
nonequilibrium steady states in our model and the associated phase
transitions are parametrized by
$n,\epsilon,q_1,q_2$.  

It is useful here to compare our model with the existing
literature on TASEP with disorder. For instance
Refs.~\cite{newzia,new}, discuss the effects of multiple defects in
an open TASEP channel (see also Ref.\cite{tasep} for general
discussions on inhomogeneous TASEPs). In the model of
Ref.~\cite{newzia} one or two point defects have been considered.
Their effects on the steady state current has been obtained. This
has been generalized in Ref.~\cite{new}, where instead of point
defects, extended defects of variable sizes are discussed.
Subsequently, Ref.~\cite{nossan} has studied the TASEP with
site-wise disorder and has obtained a set of exact results in the
low current regime. In contrast to the models in
Refs.~\cite{newzia,new,tasep,nossan}, our model is a closed model
having no edge or boundary effects, and thus with no entry or exit
of particles. Evidently, the dynamics  keeps the particle number in
our model strictly conserved. Conservation laws are known to affect
the universal scaling properties of fluctuations in equilibrium or
nonequilibrium systems~\cite{halperin,fns}. How conservation laws
affect the ensuing (possibly nonuniform) steady states in
inhomogeneous nonequilibrium systems remains a theoretically
interesting question. Our model is ideally suited to study this
issue. In particular, the two defect sites in our model in general
have unequal hopping rates ($q_1\neq q_2$), unlike the models in
Refs.~\cite{newzia,new}, where the bottlenecks are considered to
have equal hopping rates.

\section{Steady state density profiles}

On general ground, the system should be in three different phases:
(i) {\em Low Density (LD)} [{\em High density (HD)}], with the
lattice being nearly empty [full] and consequently the bottlenecks
affecting the density profile only locally in the
form of a boundary layer (BL) of vanishing thickness in TL behind the bottleneck, 
and (ii) {\em Intermediate Density (ID)}, with $n$ between LD and HD
phases, when there are macroscopic effects of the defects on the
density profile in the form of generic LDWs and their delocalization
transitions. Notice that the dynamics of TASEP is formally given by
rate equations for every site, that involves nonlinear coupling with
neighboring sites, and hence not closed~\cite{tasep}.  In our work
below, we use analytical mean-field theory (MFT), complemented by
our extensive Monte-Carlo Simulations (MCS)
(with random sequential updates), 
 for quantitative descriptions of these steady-states. In
 MFT descriptions, one proceeds
by neglecting spatial correlations and replacing {\em products of
averages} by {\em averages of products}. We begin with
 the analysis for the nontrivial ID phase below, followed by the LD
 and HD phases.

\subsection{Intermediate density phase}\label{idphase}

\begin{figure}[htb]
\includegraphics[height=6.0cm]{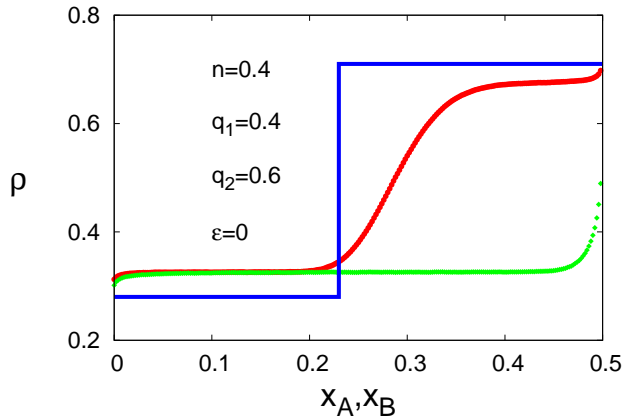}
\caption{(color online) MFT (blue continuous line) and MCS (points) plots: LDW
in $T_A$, LD in $T_B$, $\rho=[\rho_A \mbox{(red circles)},\rho_B \mbox{(green rhombus, nearly flat)}]$.}\label{ldw}
\end{figure}

We study  inhomogeneous steady states for moderate densities ({\em
Intermediate density (ID)} phase) by using analytical mean-field
theory (MFT), developed by exploiting the spatial constancy of the
steady-state currents, complemented by  extensive Monte-Carlo
Simulations (MCS)
(with random sequential updates). 
The steady state density $\rho (x)$ follows
\begin{equation}
(2\rho-1)\partial_x\rho=0,\label{steadyst}
\end{equation}
 neglecting $O(1/N)$ terms
which are insignificant in the bulk in TL (strictly, Eq.~(\ref{steadyst}) 
holds away from the BLs, which may form close to a defect). Thus, in the bulk, $\rho$
should be a constant~\cite{const1}. Therefore, in ID phase, $\rho$
can be piecewise continuous without any spatial variation, with the
possibility of an LDW in the system. The system can be viewed as a combination of
{\em two TASEP channels} $T_A (0\leq x\leq (1-\epsilon)/2)$, 
marked as a blue dashed-dotted line in Fig.~\ref{modeldiag}, and $T_B
((1-\epsilon)/2\leq x\leq 1)$  (black dashed line in Fig.~\ref{modeldiag}), 
joined at $x=0$ and $x=(1-\epsilon)/2$
respectively~\cite{erwindefect}; see Fig.~\ref{modeldiag}. 
Channels $T_A$ and $T_B$ are generally of unequal length. 
Define $x_A=(1-\epsilon)/2 -x,\,0\leq x\leq (1-\epsilon)/2,\,\,x_B=
1 -x,\,(1-\epsilon)/2 <x<1$ with $\rho_p(x_p)$ as densities for
$T_p\,(p=A,B)$. Thus, in terms of $x_A$ and $x_B$, locations of
$q_1$ and $q_2$ are given by $x_A=(1-\epsilon)/2$ and
$x_B=(1+\epsilon)/2$, respectively. We establish below the
conditions on $q_1,q_2,n$ for ID phase. If for both $q_1,q_2$, ID
phase holds (see below), there should be an LDW for each of them.
Since the LDW height depends on the hopping rates at the bottleneck,
the two putative LDWs due to $q_1,q_2$ impose {\em different steady
state currents} in different bulk regions of the system, which is
unphysical. Assuming the principle of minimum
current~\cite{min-curr}, $min(q_1,\,q_2)$ that imposes the minimum
current in the system creates an LDW behind it; the other bottleneck
creates only a boundary layer (BL) with a vanishing thickness in TL,
being rendered {\em subdominant} or {\em irrelevant} in TL (see
Fig.~\ref{ldw}).

\begin{figure}[htb]
\includegraphics[height=6cm]{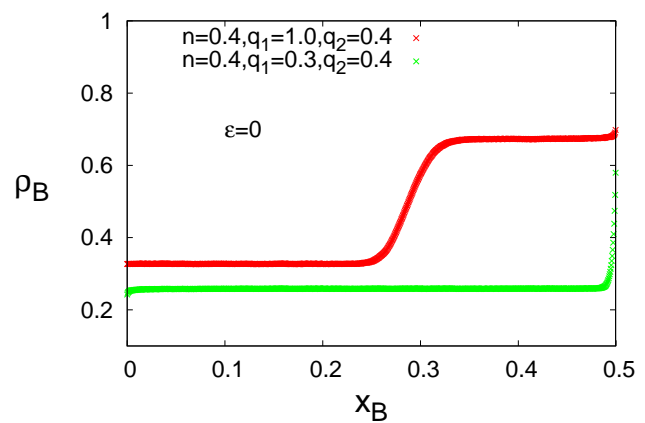}
\caption{(color online) MCS results on the screening of  $q_2$ by
$q_1$; LDW due to $q_2$ for $q_1=1$ [red (dark gray)], BL at $x=1/2$ due to
$q_2$  for $q_1<q_2$ [green (light gray)].} \label{screen}
\end{figure}

For concreteness, now consider  $q_1<q_2$ and
 an LDW in $T_A$ at $x^w_A$;  $T_B$  has a uniform density $\rho_B(x_B)=\rho_2$ in the
 bulk. Assume $n\leq 1/2$ ( $n\geq 1/2$ may be analyzed
 by the particle-hole symmetry).
 Then,
\begin{equation}
 \rho_A(x_A)=\rho_3 + \theta (x_A-x^w_A)(\rho_1-\rho_3),\,({\rm with}\, \rho_1\neq\rho_3).
 \end{equation}
 In addition, at $x=(1-\epsilon)/2$, $\rho_B(x)$ has a BL of value $\tilde{\rho_2}$. Current
 conservation at $x=0$ leads to $\rho_1(1-\rho_1)=q_1\rho_1
 (1-\rho_2)=\rho_2(1-\rho_2)$ yielding
 \begin{equation}
 \rho_1=1/(1+q_1),\,\rho_2=q_1/(1+q_1).
 \end{equation}
  Further, current
 conservation in $T_A$ yields
 $
\rho_1(1-\rho_1)=\rho_3(1-\rho_3)$ and since $\rho_1\neq \rho_3$,
\begin{equation}
\rho_3=1-\rho_1=1-1/(1+q_1)=q_1/(1+q_1).
\end{equation}
 For a BL of
height $\tilde{\rho_2}$ and vanishing thickness in TL at $i=N$,
current conservation leads to
\bea
q_2\tilde{\rho_2}(1-\rho_3) =
\rho_3(1-\rho_3)  \Rightarrow \tilde{\rho_2} = {1 \over q_2}\rho_3.
\eea
From current conservation at $x=0$, $\rho_2={q_1 \over
1+q_1}=\rho_3$. Further, particle number conservation (PNC)
\begin{equation}
\int_0^{(1-\epsilon)/2} dx_A \rho_A(x) +
\int_0^{(1+\epsilon)/2}\rho_B(x_B)dx_B=n
\end{equation}
 yields (see Fig.~\ref{ldw})  
 \bea x^w_{A}= {1+q_1
\over
1-q_1}\left(\frac{1}{2}-\frac{1-q_1}{2(1+q_1)}\epsilon-n\right),
\label{dw}
 \eea
  as the LDW position in $T_A$. Equation (\ref{dw}) appears to yield a diverging
$x^w_A$ as $q_1\rightarrow 1$. However, in that limit with
$q_2>q_1$, the model is no longer in the ID phase and Eq.~(\ref{dw})
does not apply. As expected, $x^w_A$ depends only on
$q_1$~\cite{ldw1}. Notice that $x_A^w=(1-\epsilon)/2$; thus an LDW
is just formed at the location of $q_1$. Hence, this yields
$q_1=n/(1-n)$, or $n=q_1/(1+q_1)= \rho_{LD}$, setting the boundary
between the ID and {\em Low Density (LD)} phase. Particle-hole
symmetry immediately yields $q_1=(1-n)/n$, or,
$n=1/(1+q_1)\equiv\rho_{HD}$ as the boundary between the ID and {\em
High Density (HD)} phases. Thus, for $\rho_{LD}< n <\rho_{HD}$, ID
phase ensues. If for some $n$, both $q_1,q_2,q_1<q_2$ satisfy ID
phase conditions, then there is an LDW due only to $q_1$; the
putative LDW due to $q_2$, that would have existed with $q_2\neq 1$
and $q_1=1$, gets suppressed to a BL with a vanishing thickness in
TL. Thus, $q_2$ gets {\em screened} or rendered irrelevant by $q_1$,
establishing an universal feature here~\cite{screen}; see Fig.~\ref{screen},
where we show an LDW due to $q_2=0.4$ with $q_1$ set to unity and
its suppression when $q_1=0.3<q_2$. Notice that this notion of
universality is distinct from its significance elsewhere, e.g., in
equilibrium critical phenomena and (equilibrium or nonequilibrium)
critical dynamics. In these latter examples, universality implies
correlations of fluctuations (generally about uniform mean
backgrounds) are independent of the model parameters. In contrast,
in the present case, the idea of universal features concerns the  mean
density profile itself; in addition, it does not imply that the
ensuing steady state density profile is independent of the model
parameters, since the LDW explicitly depends upon (in addition to
$n$) the strength of the strongest bottleneck ($q_1$ in the above
example), which is a microscopic model parameter. It may be
noted that the height of an LDW is determined by the current
conservation across the dominant defect, whereas its position is
determined by PNC.


\begin{figure}[htb]
\includegraphics[height=6.0cm]{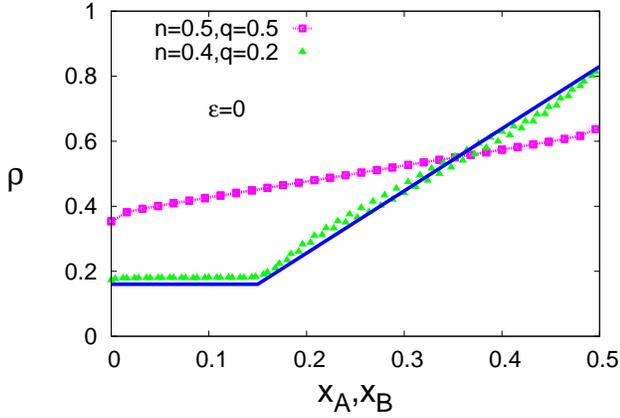}
\caption{(color online)  Overlapping DDWs in $T_A$ and $T_B$; blue
continuous line represents MFT results for $n=0.4, q=0.2$. Note the
agreement between MCS (green triangles) and MFT results.}\label{ddwoverlap}
\end{figure}


What happens when $q_1=q_2=q$? Then, the conditions for LDWs
due to $q_1,\,q_2$ are the same; hence two LDWs, one each in $T_A$
and $T_B$, should form.
 With $x^w_A$ and $x^w_B$ as the locations of the LDWs  in
$T_A$ and $T_B$, respectively, PNC yields a linear relation between
them, without determining them uniquely, and hence, we obtain one
DDW in each of $T_A$ and $T_B$. This can be physically understood as
a consequence of PNC: If there are  two LDWs at $x^w_A,\,x^w_B$ in
the system (since $q_1=q_2$, both satisfying the ID phase
conditions), PNC obviously holds true by shifting $x^w_A,\,x^w_B$ by
equal and opposite amounts, resulting into arbitrariness in the
values of $x^w_A,\,x^w_B$. This manifests into two DDWs, one each in
$T_A$ and $T_B$. Long time averaged $\rho_A(x_A)$ and $\rho_B(x_B)$,
unlike an LDW, do not display any discontinuity, instead take the
form of
 inclined lines, representing the envelops of the two DDWs (Fig.~\ref{ddwoverlap}). For $\epsilon=0$ an
estimation of $\Delta$, the span of each DDW may be made. Notice
that PNC together with $q_1=q_2$ dictates that under long-time
averaging $\rho_A(x_A)=\rho_B(x_B)$; we write for the {\em average
locations} of the DWs $\langle x^w_A\rangle = \langle x^w_B\rangle
=x_0$, where $\langle..\rangle$ represents averages over steady
state realizations.  Assuming a linear profile for the DDWs
(consistent with our MCS  data),  PNC then yields~\cite{ddwpos} for
the mean position of  the DDW $x_0$:
\begin{equation}
x_0 = -\frac{1+q}{2(1-q)}[n-\frac{1}{1+q}].\label{ddwposi}
\end{equation}
Since particles accumulate behind the bottleneck(s), each DDW
wanders a distance $\Delta (\epsilon=0)=2(1/2-x_0)$, allowing us to
reconstruct the DDW profiles. Equation~(\ref{ddwposi}) gives
$\Delta/2=1/2-x_0=1/4$ for all $q<1$ and $n=1/2$, corresponding to
DDWs spanning $T_A$ and $T_B$ entirely. For all other $n$, the span
is generally smaller; see Fig.~\ref{screen} showing DDWs (from MC
and MFT) for $n=1/2,0.4$, in agreement with our analysis here.

\begin{figure}
\centering
\includegraphics[height=5.5cm,width=9cm]{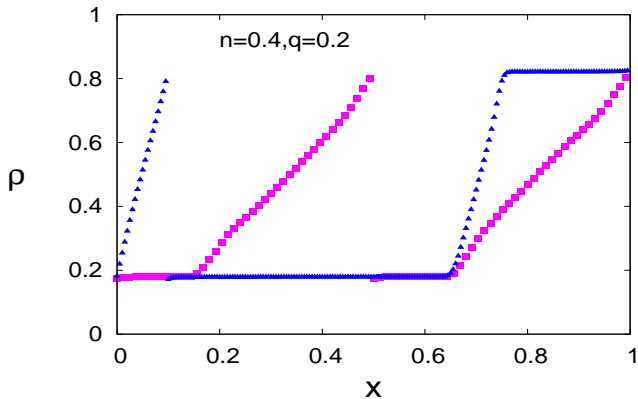}
\caption{(color online) DDW profiles in $\rho_A$ and $\rho_B$ for various
$\epsilon$.  $\epsilon=0$: $\rho_A$ (pink square left) and $\rho_B$ (pink square right);
$\epsilon=0.8$: $\rho_A$ (blue triangle left) and $\rho_B$ (blue triangle right). Changes in
the DDW spans are clearly visible.} \label{ddwtot}
\end{figure}

\begin{figure}[htb]
\includegraphics[height=6cm,width=7.5cm]{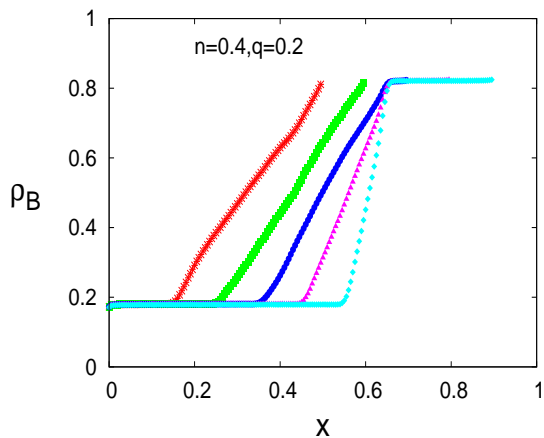}
\caption{(color online)  DDW confinement in $T_B$ for increasing
$\epsilon:$\; (from left to right) $\epsilon=0$({red}), 0.2({green}), 0.4({blue}),
0.6({magenta}), 0.8({cyan}); no change in $\Delta(\epsilon)$ from
$\Delta (\epsilon=0)$ for $\epsilon\leq \epsilon_c=0.3$ (MFT result;
not shown in the Fig.)}\label{ddwconf}
\end{figure}

\begin{figure}[htb]
\includegraphics[height=6cm,width=7.5cm]{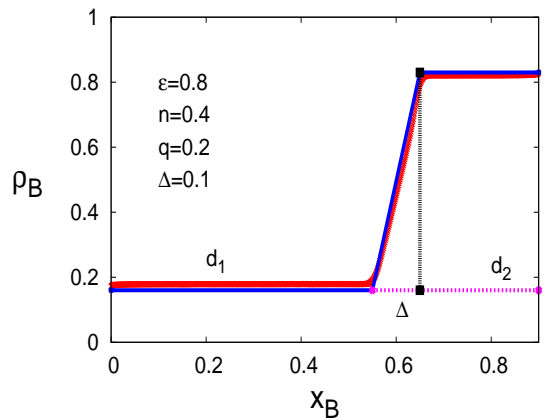}
\caption{(color online)DDW in $T_B$ for $\epsilon>\epsilon_c$; good
agreement between MFT (blue continuous line segments) and MCS (points) shown.}
\label{ddwepsilon}
\end{figure}

\begin{figure}
\includegraphics[height=6cm,width=7.5cm]{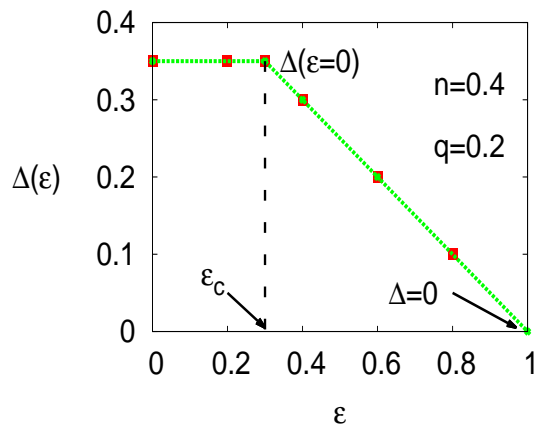}
\caption{(color onlne) Variation of $\Delta(\epsilon)$ with
$\epsilon$ with complete DDW confinement for $\epsilon=1$ from MFT
(dotted line) and MCS (points) results. } \label{ddwfig}
\end{figure}

For $|\epsilon|\neq 0$,  $\rho_A$ and $\rho_B$ are no longer
identical under long time averaging.  However, the DDW spans remain
equal in $T_A$ and $T_B$ due to PNC. We now heuristically obtain the
DDW profiles. Noting that the particles tend to accumulate right
behind the bottlenecks, as long as $\Delta (\epsilon=0) <
(1-\epsilon)/2$, DDW excursions are {\em not} expected to be
affected by shortening of $T_A$ ($\epsilon>0$) at the simplest level
of description. Hence, when
\begin{equation}
\Delta(\epsilon=0)<(1-\epsilon)/2,
 \end{equation} we set
\begin{equation}
 \Delta(\epsilon\neq 0)=\Delta (\epsilon=0)
\end{equation}
 for both $\rho_A$ and $\rho_B$.  This, together with PNC,
yields the full DDW profiles in $T_A$ and $T_B$. For
\begin{equation}
\Delta (\epsilon=0)\geq (1-\epsilon)/2,
\end{equation}
 assuming that $T_A$ and
$T_B$ may still be treated as two different TASEPs, DDW in $T_A$ is
expected to be fully contained in it,
\begin{equation}
\Delta(\epsilon\neq 0)=(1-\epsilon)/2
\end{equation}
 for $\rho_A$
and hence for $\rho_B$ as well. Full profile of $\rho_A$ is obtained
trivially. PNC and $\Delta(\epsilon)$ together then yield $\rho_B
(x_B)$: Assume that $\rho_B(x_B)$ has a low density part of length
$d_1$, a high density part of length $d_2$ and a DDW part of length
$\Delta(\epsilon)$, as shown in Fig.~\ref{ddwepsilon}, such that
\bea d_1 +d_2+\Delta(\epsilon)=(1+\epsilon)/2. \label{d1d2} \eea By
using PNC we obtain
 \bea &&\int_0^{(1-\epsilon)/2}\rho_A (x_A) dx_A
+\int_0^{(1+\epsilon)/2}\rho_B(x_B) dx_B=n,
\nonumber \\
&\Rightarrow& \int_0^{(1+\epsilon)/2}\rho_B(x_B) dx_B=n -
\int_0^{(1-\epsilon)/2}\rho_A (x_A) dx_A,
\nonumber \\
&\Rightarrow& d_1{q \over 1+q}+\Delta{q \over 1+q}+{\Delta \over
2}{1-q \over 1+q} +
d_2{1 \over 1+q}\nonumber \\&=&n- \Delta{q \over 1+q}-{\Delta \over 2}{1-q \over 1+q} , \nonumber \\
&\Rightarrow& d_1q+d_2=(n-\Delta)(1+q), \label{d1qd2} \eea and
Eq.~(\ref{d1d2}) together then yield $\rho_B (x_B)$ in terms of the
parameters $d_1,d_2$ and $\Delta (\epsilon)$. The values of $d_1$
and $d_2$ are found to be
 \begin{eqnarray}d_1&=&\frac{1+\epsilon}{2(1-q)}
+\frac{q\Delta }{1-q} - \frac{n(1+q)}{(1-q)},\\
d_2&=&-\frac{(1+\epsilon)q}{2(1-q)} - \frac{\Delta}{1-q} +
\frac{n(1+q)}{1-q}.\end{eqnarray}
  Now, define a critical $\epsilon_c$ by
$\Delta(\epsilon=0)=(1-\epsilon_c)/2$, such that for $\epsilon\geq
\epsilon_c$, $\Delta(\epsilon=0)\geq (1-\epsilon)/2$;
$\Delta(\epsilon)$ decreases linearly with $\epsilon$, reducing to
zero for $\epsilon=1$ for which $T_A$ effectively shrinks to a
point. Hence, $\Delta$, the DDW  span in $T_B$,  gets {\em
 shortened} with increasing $\epsilon$, thus confining
 DDW in $T_B$, eventually reducing to zero for $\epsilon=1$, corresponding to an
LDW in $T_B$ for $\epsilon=1$ in TL; the corresponding DDW in $T_A$,
which becomes the coincident location of the two defects (of equal
magnitude) for $\epsilon=1$, naturally reduces to an LDW at the
coincident point of the two defects~\cite{lebo1}. This establishes
confinement of DDWs in our model. Our MF analysis for DDWs are
complemented by MCS studies: See Fig.~\ref{ddwtot} for DDW profiles
for $\rho_A$ and $\rho_B$
 for $\epsilon=0$ and $\epsilon=0.8>\epsilon_c=0.3$ (MFT value).
 Figure~\ref{ddwtot} in fact clearly illustrates sharpening (i.e.,
 confinement) of the DDW profiles of both $\rho_A$ and $\rho_B$.
 See also Fig.~\ref{ddwconf} for the DDW profile
 for $\rho_B(x_B)$ for various values of $\epsilon$ and Fig.~\ref{ddwepsilon} for the agreement
between MFT and MCS result which clearly confirms our intuitive
arguments above~\cite{ddw1}. A plot of $\Delta(\epsilon)$ versus
$\epsilon$ is given in Fig.~\ref{ddwfig}.

The phenomenon of confinement may be understood heuristically as
follows. Notice that in the ID phase, the heights of the HD and LD
parts of an LDW due to the dominant defect are entirely determined
by current conservation at the defect and are independent of
$\epsilon$. Since DDWs essentially are long-time averages over LDW
profiles, all of which have the same heights for their LD and HD
parts, the values of the densities at the lowest and highest points
of the corresponding DDW envelope should be independent of
$\epsilon$ as well. Furthermore, the DDW span $\Delta$ being
determined by PNC, it remains unaffected by changes in $\epsilon$ so
long as $\Delta (\epsilon=0)< (1-\epsilon)/2$. Beyond
$\epsilon=\epsilon_c$, i.e., $\Delta(\epsilon=0)=(1-\epsilon)/2$,
qualitatively speaking there exists two different possibilities for
$\Delta(\epsilon\neq 0)$: either $\Delta(\epsilon\neq 0) >
(1-\epsilon)/2$, or, $\Delta(\epsilon\neq 0)=(1-\epsilon)/2$. If the
former case holds, the defect will be {\em inside} the DDW envelop.
This should imply (a) there should be more particles on average in
front of the defect than behind, and (b) consequently, current
conservation across the defect will be violated. Since these
possibilities are unacceptable, we discard option (a) and set
$\Delta(\epsilon\neq 0) = (1-\epsilon)/2$ for $\epsilon\geq
\epsilon_c$, which evidently satisfies current conservation at the
defect(s) for all the individual LDW profiles making up a particular
DDW envelope. Our heuristic arguments are clearly validated by our
MCS simulation, as displayed in Fig.~\ref{ddwconf}. Notice that
TASEPs with open boundaries also exhibit DDWs for equal entry and
exit rates, both being less than 1/2~\cite{rev2}. This is due to the
uncorrelated entry and exit of particles in an open TASEP. The span
of a DDW in an open TASEP covers the entire system. In contrast,
DDWs here are due to the indeterminacy of the corresponding LDW
positions subject to PNC, or equivalently, the freedom in fixing the
LDW positions while maintaining PNC. Furthermore, the span of the
DDWs in the present model is determined by PNC, along with
$\epsilon$. Thus, PNC plays a crucial role in DDW formation in the
present model, unlike open TASEPs.

\subsection{LD and HD phases}\label{ldhd}

\begin{figure}[htb]
\includegraphics[height=5.5cm]{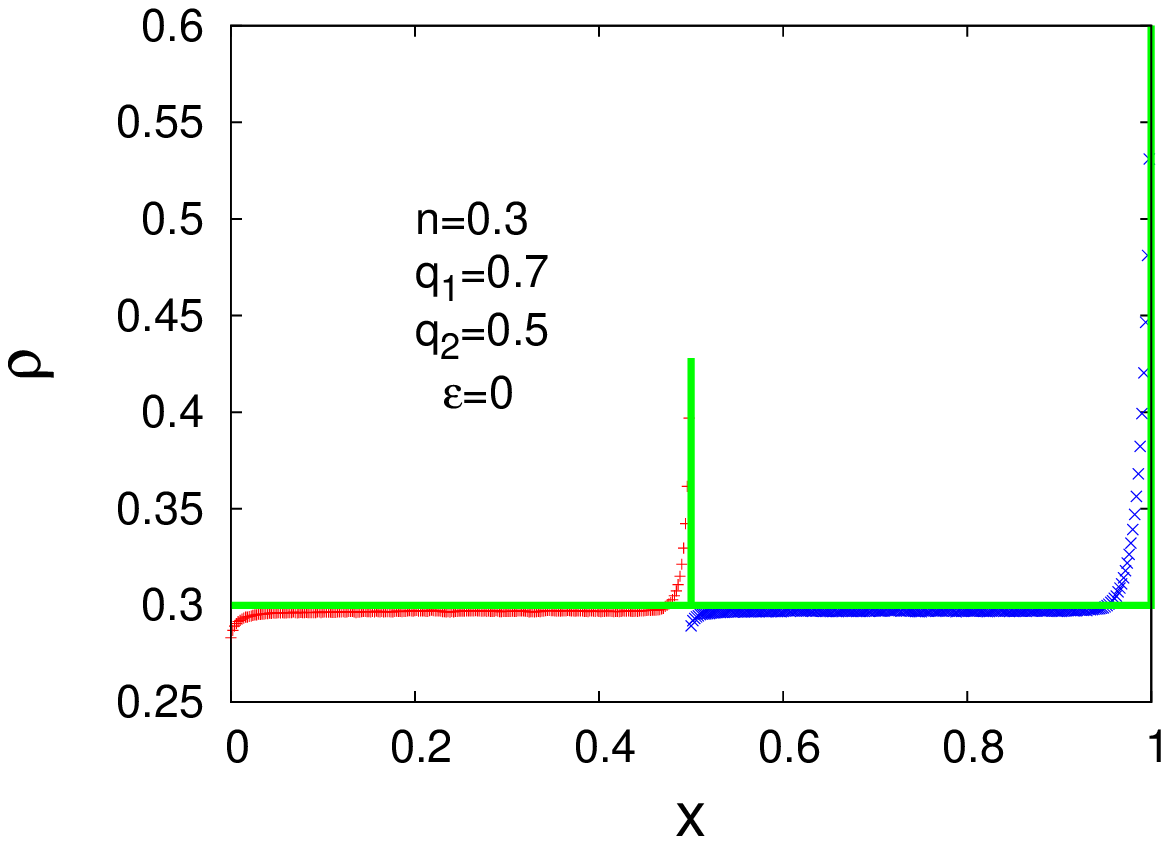}\\
\includegraphics[height=5.5cm]{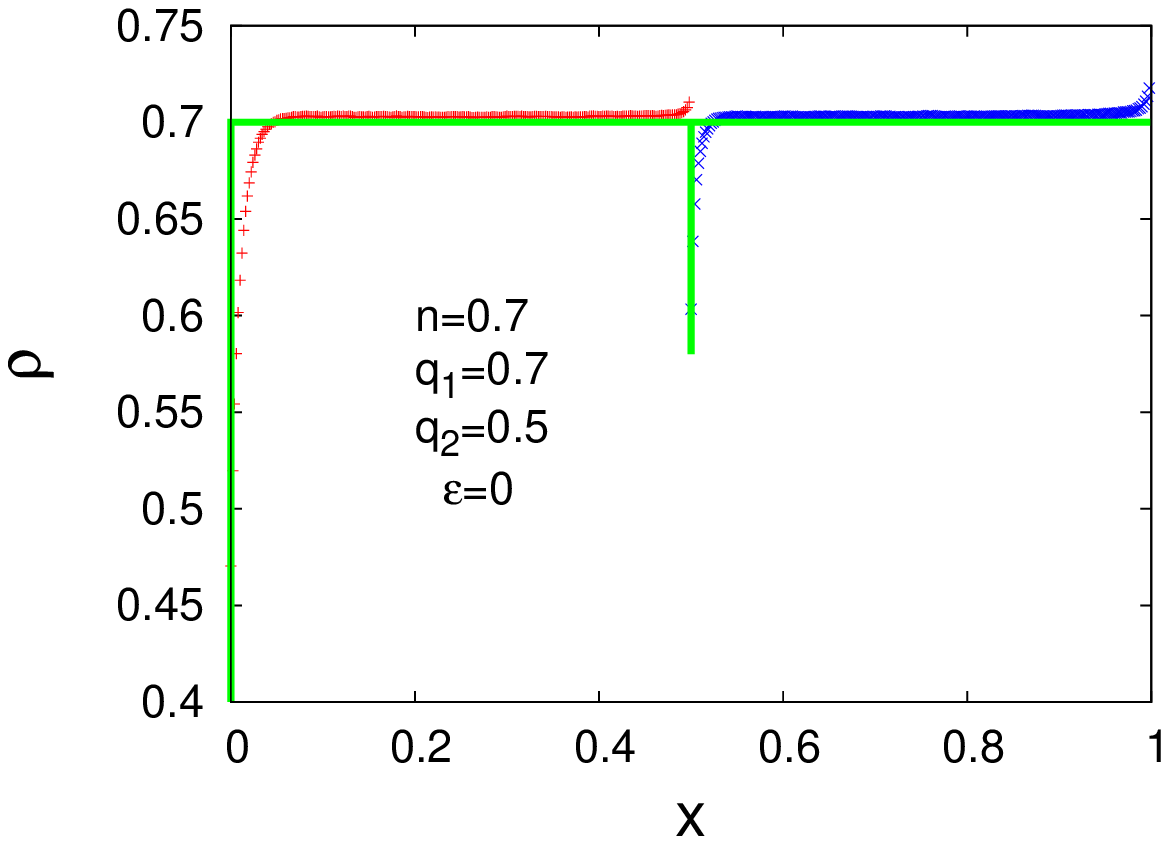}\caption{(color online) LD and HD phases. MFT (green continuous lines)
and MC (points) are shown.} \label{lhd}
\end{figure}

Consider now the LD phase: the system is diluted and the particles
are well separated. In such a {\em low density traffic} situation,
we do expect
 the bulk density in TL should be same as the overall mean
density, $n=N_p/(2N)$. Just a local peak (a BL) in the density at
the bottleneck with a vanishing thickness in TL appears, such that
$\rho=n +h_m$, $m=N(1-\epsilon),2N$, $h_m$ being the local jump
height imposed by the defects at $i=m$. Thus using MFT in TL,
current conservation yields $h_m=n(1-q_m)/q_m$,  $q_m=q_1,q_2$.
Hence, as $q_{1,2}\rightarrow 0$, i.e., as the bottlenecks grows
stronger, i.e., $q$ decreases, the
peak height $h_m$ grows bigger and current $j$ decreases. 
Now if this decrease in  $j$ is large enough 
such that $j\leq j_c$, a threshold critical value,
the bottlenecks starts to have global or macroscopic effect on the
system. We minimize $j$ to get a maximum critical value of $h_m$ and
thence a critical density~\cite{erwindefect}
\bea
\rho_{LD,m}={q_m \over 1+q_m},
\eea
such that the
LD phase prevails so long as $n<\rho_{LD,m}$, beyond which 
 the bottlenecks have macroscopic effects. The HD phase of the
system may be analyzed by using the particle-hole symmetry yielding
a critical density
\bea
\rho_{HD,m}=({1 \over 1+q_m}),
\eea
such that for
$n<\rho_{HD,m}$ the macroscopic effects of the bottlenecks manifest,
else, the bottlenecks impose only BLs (as local dips)  in the
density having vanishing width in TL. Thus, in both LD and HD
phases, $\rho$ is independent of the bottlenecks in TL.  Plots of
the density profiles in the LD and HD phases are shown in
Fig.~\ref{lhd}.
As soon as $n > min( \rho_{LD,m})$ or $n < max(\rho_{HD,m})$, the
effects of the bottlenecks are no-longer
localized~\cite{erwindefect} and the system is in its ID phase.
Unsurprisingly, these conditions are identical with the conditions
for the threshold of the ID phase derived above independently.
Overall, then  LD and HD phases are characterized by macroscopically
uniform densities $n=N_p/(2N)$, with BLs of vanishing thicknesses in
TL forming behind $q_1,\,q_2$. Thus coarse-grained density
measurements will not detect any of $q_1,q_2$ in LD/HD phases.

\section{Phase diagram}\label{phasediag}

\begin{figure}[htb]
\includegraphics[height=6.5cm]{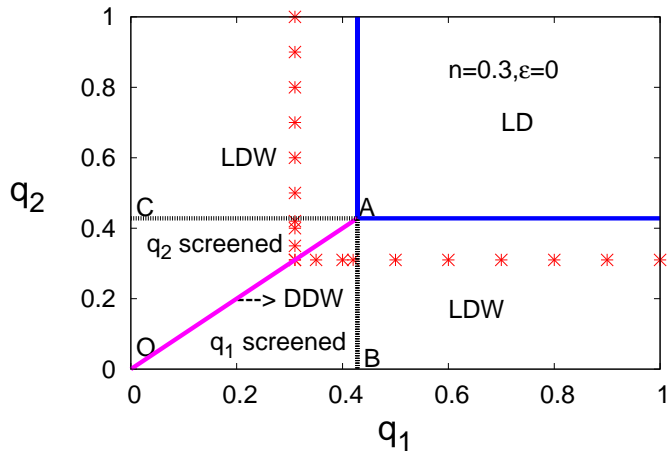}
\caption{ (color online) MFT (lines) and MCS (points) phase diagram
in $q_1-q_2$ plane. DDW (line AO), screening of $q_1$ by $q_2$
(triangle AOB) and vice versa (triangle AOC) shown.}
\label{phaseplot}
\end{figure}

Let us now consider the phase diagram of the model in the $q_1-q_2$ plane for a fixed $n$.
Consider $n<1/2,\epsilon=0$. Thus, $n<1/(1+q)$ is {\em
always} satisfied, since $q\leq 1$. Therefore, the system can be
only in LD or ID phases, but not in the HD phase for $n<1/2$. For
$q_1,q_2>n/(1-n)$, the LD phase follows, else  the ID phase ensues
with
 a pair of DDWs are found along the line $q_1=q_2=q\leq n/(1-n)$. For $n>1/2$ similar arguments follow, leading to the system showing
only the HD or ID phase with no LD phase possible. The corresponding
phase boundaries may be similarly obtained phase. This is consistent
with the particle-hole symmetry of the model.
 For half filling
($n=1/2$), both $q/(1+q)=1/2$ and $1/(1+q)=1/2$ yields  $q=1$, so
that for $(q_1,q_2)\leq 1$, the ID phase prevails with DDWs along $
q_1=q_2$, with no LD/HD phases.
 More generally, as $n\rightarrow 1/2$, the area covered by the ID phase in the phase diagram increases,
covering the entire phase diagram for $n=1/2$. Evidently, MFT and
MCS results, though agree qualitatively, lack quantitative
agreement, presumably due to the correlation effects neglected in
MFT (see Fig.~\ref{screen}(a) above for equivalent quantitative
disagreements between MFT and MCS results for density profiles). We
have used various system sizes in our MCS studies, ranging from
$2N=500$ to 2000, all of which agree with each other within the
numerical accuracies of our MCS studies, ruling out any significant
system size effects.  Notice that in region AOC (AOB) of
Fig.~\ref{phaseplot}, both $q_1,q_2$ satisfy ID phase condition, but
$q_2(q_1)$ is screened by $q_1(q_2)$. Now with the current 
$J_m=\rho_m(1-\rho_m),\,m=A,B$ for channels $T_m$ in the bulk, $J_m$
is clearly continuous across the phase boundaries in Fig.~\ref{phaseplot}, since density $\rho_m$
changes continuously across the phase boundaries. This is
reminiscent of a second order phase transition between the LD and ID phases (and hence between
the HD and ID phases by using the particle-hole symmetry in the model). 
Equivalently, considering the LDW position as an
order parameter, the phase boundaries in Fig.~\ref{phaseplot} are
second order in nature, with an order parameter exponent 1. This
may be obtained as follows: Assume $q_1<q_2$ (thus $q_2$
irrelevant). Then,
 to obtain the behavior of $x^w_A$ near the LD-ID phase transition, use Eq.~(\ref{dw}) and set
\begin{equation}q_1=q_c-\delta q,\delta q>0,\end{equation}
 with $q_c=n/(1-n)$ at the
threshold of the ID phase for a given $n$. This yields,
\begin{equation}
x_A^w={(1-\epsilon) \over 2}-{\delta q (1-2n+\epsilon) \over
2(1-q_c)},\end{equation} for small $\delta q>0$, where we have used
Eq.~(\ref{dw} to obtain the above. Now, define an order parameter
$O=x_A^w-(1-\epsilon)/2$, such that it is zero in the LD phase and
non-zero in the ID phase. For small $\delta q>0$, then, $O=\delta
q{(2n-1-\epsilon) \over 2(1-q_c)}$, giving the order parameter
exponent 1, with $q_1$ as the control parameter (analog of
"temperature" in an equilibrium phase transition). This is in
contrast to a typical mean-field value of 1/2 for the order
parameter exponents in equilibrium systems~\cite{expovalues}.

\section{Conclusion and outlook}\label{conclu}

 This work, thus,  shows how the mutual interplay between PNC
  and bottleneck competition leads to a complex macroscopic behavior
including delocalization transitions of LDWs and confinement of
DDWs. While,  coarse-grained measurements of an LDW (or DDWs) reveal
the strength and (relative) position(s) of the strongest
bottleneck(s), but screening prevents detecting the subdominant
bottlenecks. Equivalently, different systems having the same
particle density and strongest bottleneck(s), but with varying
(in number, strength and relative positions) subdominant
bottlenecks yield same macroscopic density profiles, revealing an
underlying universal feature.  This universal feature uncovered here has strong
experimental implications. For instance, ribosome density
mapping~\cite{riboden} or ribosome density profiling~\cite{riboprof}
experiments  measuring coarse-grained densities can detect only the
strongest pause sites (or non-preferred codons), but cannot resolve
the other weaker (subdominant) pause sites. As we have already
discussed above, the notion of universality in the present context
is conceptually distinct from its implications elsewhere. We also
take note of the fact that in our model particle number conservation
affects the resulting macroscopic density profiles in ways that are
very different from its role in other systems, e.g., universal
critical dynamics in equilibrium or nonequilibrium systems.
Phenomenologically, DDW confinements imply that tuning the
bottleneck positions, it is possible to control the extent of
movement of inhomogeneous densities, a feature expected to be
significant for {\em in-vitro} set ups. Our results may be tested in
{\em in-vitro} experiments by studying the restricted 1D motion of
micron-sized self-propelled (active) particles along circular rings
with constrictions~\cite{active}. In addition, general features of
our results should be observed in vehicular jams in a closed network
of roads with bottlenecks, e.g., in Formula 1 tracks where car
speeds are reduced near the sharp bends ("bottleneck"), resulting
into accumulation of cars behind them~\cite{F1}. We now make a brief
comparison between our results and those of Refs.~\cite{newzia,new}.
The latter works typically found localized shocks or LDW. In
addition, Ref.~\cite{new} also found that a second, smaller
bottleneck, far form the first one has no effect on the current. In
particular in both Refs.~\cite{newzia,new} the hopping rate across
the defects (point of extended) have the same magnitude. Our closed
model, in contrast, display DDWs in addition to LDWs, the associated
localization-delocalization transitions and confinement of the DDWs.
Furthermore, screening of the weaker defect in our model can happen
regardless of the mutual distance between the weaker and the
stronger defects. Note that in general we have unequal hopping rates
at the point defects, unlike the models in Refs.~\cite{newzia,new}.

 Our work is a promising starting point for understanding systems
with a large number of discrete bottlenecks. 
For intermediate values of $n$, macroscopically inhomogeneous
density profiles ensue. With nonidentical bottlenecks, the strongest
one, (i.e., with the lowest hopping rate) controls the macroscopic
inhomogeneity in the form of an LDW, whose position may be obtained
by above analysis together with screening of the weaker bottlenecks.
 When there are more than one strongest bottleneck, those many
DDWs will be formed, as here. 
Nonetheless, screening of weaker bottlenecks and its experimental
implications should generally hold. Our model is complementary to
the model in Ref.~\cite{must}. It will be interesting to study how
density profiles for discrete, isolated bottlenecks are modified
eventually reducing to the results in Ref.~\cite{must}. Lastly,
considering the central role of number conservation in the present
model, it will be interesting to see how violations, especially weak
violations of particle number conservation affect the steady states
in this system~\cite{nilu1}. 



\section{Acknowledgement} One of the authors (AB) wishes to thank the
Max-Planck-Gesellschaft (Germany) and Department of Science and
Technology/Indo-German Science and Technology Centre (India) for
partial financial support through the Partner Group programme
(2009). NS would like to thank S. Biswas, A. Ghosh and A. Chandra
for fruitful discussions.

\appendix

\section{Mean-field Phase diagram in the $n-q_1$ plane}

\begin{figure}[h]
\includegraphics[height=7cm,width=7.5cm]{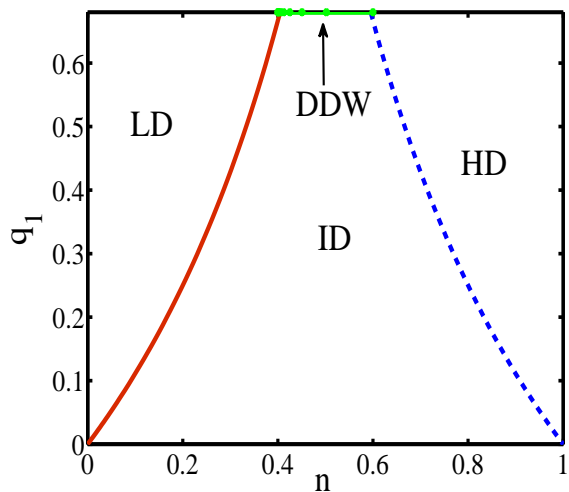}
\caption{ (color online) Mean-field phase diagram in $n-q_1$ plane
($q_2=0.68$). LD, ID and HD phases and the DDW line are shown. Note
that the limit of $q_1$ is confined between 0 and $q_2=0.68$.}
\label{phase2}
\end{figure}

An analogous mean-field phase diagram in the $n-q_1$ ($q_1\leq
q_2,\,\epsilon=0$) plane is shown in Fig.~\ref{phase2}. The LD, HD
and ID phases are shown; $q_1=n/(1-n)$ gives the boundary between
the LD and ID phase; $q_1=(1-n)/n$ gives the boundary between the ID
and HD phases. The upper limit of $q_1$ is confined up to $q_2$,
since above this value, $q_1$ gets screened by $q_2$; $0.4<n<0.6$
gives the location of DDWs. Not surprisingly, 
the particle current is continuous across the phase boundaries in Fig.~\ref{phase2}, similar to the
continuity of the particle current across the
phase boundaries in Fig.~\ref{phaseplot}. This is consistent with the second
order phase transitions between the LD (HD) and the ID phases in the system.

\section{Locations of LDW and DDW in the $\epsilon-\Delta_q$ plane}
Consider the locations of LDW and DDWs in the $\epsilon-\Delta_q$
plane, where $\Delta_q = |q_1-q_2|$ with $n=1/2$. Evidently, the
$\Delta_q=0$ line corresponds to DDWs in the model, with DDW spans
shrinking as $\epsilon$ rises to 1, finally being fully confined at
$\epsilon=1$. The rest of the box with $\Delta_q >0$ corresponds to
LDWs in the system.
\begin{figure}[h]
\includegraphics[height=7cm,width=7.5cm]{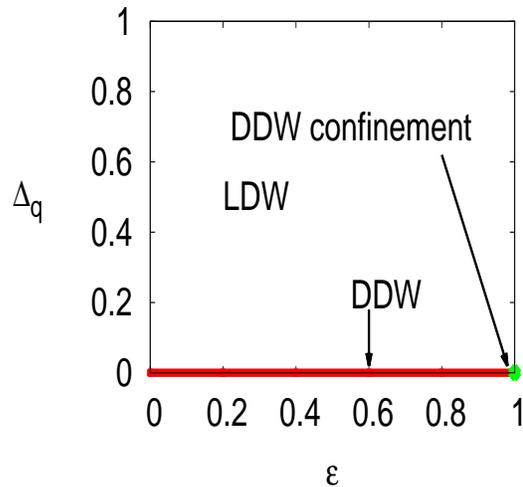}
\caption{ (color online) Locations of LDW, DDW and DDW confinement
in $\Delta_q-\epsilon$ plane, $n=0.5$.} \label{phase3}
\end{figure}


\begin{thebibliography}{99}
\bibitem{const1} TASEP with open boundaries does exhibit
non-constant steady state density with equal entry and exit rates,
both being less than 1/2. This has an analogue in the present model
(see below).
\bibitem{min-curr} This is reasonable, for in the limit when any of
$q_1,\,q_2\rightarrow 0$, the system current must vanish. Thus,
$min(q_1,\,q_2)$ controls the bulk current in the ID phase of the
system.
\bibitem{ldw1} Interestingly,  with
$n=\frac{1}{2}-\frac{1-q_1}{2(1+q_1)}\epsilon\equiv n_c$, $x^w_A=0$,
i.e., the LDW is pushed to the the location of $q_2$ at
$x=(1-\epsilon)/2$  For $n>n_c$, $x^w_A<0, |x^w_A|>(1-\epsilon)/2$,
pushing the LDW in $T_B$. This breaks down the basic MF assumption
of splitting the system into $T_A$ and $T_B$. However, an equivalent
MF analysis may be constructed by considering the hole density $n_h
= 1-n$. Since the holes move clockwise, an LDW will be formed that
is entirely contained in $T_B$ and $T_A$ will be uniform with an LD
phase for the holes. The MF hole density profiles may be found by
following the MF analysis outlined above;  the particle-hole
symmetry then directly yields $x^w_A$.
\bibitem{screen} It may be noted that the use of the term
"screening" does not imply any collective effect here. It only
refers to experimental invisibility (in a coarse-grained sense) of
the weaker bottleneck in the presence of the stronger one.
\bibitem{ddwpos} This is equivalent to  hypothetically replacing the actual
DDW density profiles in $T_A$ and $T_B$ by two identical
$\theta$-functions at $x_0$: $\rho_{A,B}=\rho_3 + \theta
(y-x_0)(\rho_1-\rho_3),\,y=x_{A,B}$.
\bibitem{ddw1} For $n>1/2$, from Eq.~(\ref{ddwposi}), $\Delta(\epsilon=0)>1/2$,
and hence, the DDW span is larger than $T_A$ or $T_B$.
 However, as discussed in Ref.~\cite{ldw1} above,  MFT may still be constructed for
the hole densities $n_h=1-n$, such that the DDW span for the holes
is smaller than $1/2$. Particle-hole symmetry then allows us to
calculate the corresponding DDW particle density profiles in the
system.
\bibitem{lebo1} In this limit, our model is identical to the one in
Ref.~\cite{lebo}.
\end{thebibliography}
\end{document}